\documentclass[aps,pra,reprint]{revtex4-1}
\usepackage{hyperref}
\usepackage{amsmath}
\usepackage{amssymb}
\usepackage[pdftex]{graphicx}
\usepackage{microtype}
\usepackage{verbatim}
\usepackage{todonotes}
\presetkeys{todonotes}{inline,backgroundcolor=yellow}{}
\usepackage{subcaption}
\usepackage{multirow}
\usepackage{array,booktabs,tabularx,tabulary}
\usepackage{natbib}
\usepackage{adjustbox}
\usepackage{placeins}
\makeatletter
\DeclareRobustCommand{\element}[1]{\@element#1\@nil}
\def\@element#1#2\@nil{%
  #1%
  \if\relax#2\relax\else\MakeLowercase{#2}\fi}
\makeatother

\newcommand{\mrm}{\mathrm}

\begin{document}
%\linenumbers
\title{Nd$^+$ isotope shift measurements in a cryogenically-cooled neutral plasma}
\author{Nishant Bhatt}
\author{Kosuke Kato}
\author{Amar C. Vutha}
\email{amar.vutha@utoronto.ca}
\affiliation{Department of Physics, University of Toronto, Toronto, Canada M5S 1A7}

\begin{abstract}

We report measurements of the isotope shifts of two transitions ($4f^46s\rightarrow [25044.7]^{\circ}_{7/2}$ and $4f^46s\rightarrow [25138.6]^{\circ}_{7/2}$) in neodymium ions (Nd$^+$) with hundredfold improved accuracy, using laser spectroscopy of a cryogenically-cooled neutral plasma. The isotope shifts were measured across a set of five spin-zero isotopes that spans a nuclear shape transition. We discuss the prospects for further improvements to the accuracy of Nd$^+$ isotope shifts using optical clock transitions, which could enable higher precision tests of King plot linearity for new physics searches.
\end{abstract}

\maketitle

\section{Introduction}

Isotope shifts (IS) of optical transitions in atoms are a valuable source of information about nuclear properties, such as nuclear charge radii and nuclear shape transitions \cite{Angeli2013,Blaum2013,Campbell2016}. Precise IS measurements are useful as experimental benchmarks against which theoretical atomic structure calculations can be compared (e.g., \cite{Kalita2018,Ohayon2019}). IS measurements have also been used to identify super-heavy elements in space \cite{Dzuba2017b}, and probe the variation of fundamental constants \cite{Kozlov2004,Murphy2013}. 

Interest in precision IS measurements has been sparked by recent proposals to use IS to probe beyond-standard-model (BSM) physics \cite{Delaunay2017,Frugiuele2017,Flambaum2018,Berengut2018}. These proposals take advantage of the fact that isotope shifts of an atomic transition can be parametrized very well by two parameters that depend on the atomic electron distribution: a field-shift coefficient that determines the sensitivity of the transition to a change in the nuclear shape and volume, and a mass-shift coefficient that determines the sensitivity to a change in the nuclear mass \cite{King1984}. Therefore the isotope shifts of two different atomic transitions, measured in the same set of isotopes, trace out a straight line when they are appropriately rescaled and plotted against each other in a King plot \cite{King1984,Blundell1987}. The nonlinearity of the King plot is then a measure of higher-order electron-nuclear interactions \cite{Yerokhin2020}, including new ones from BSM physics \cite{Delaunay2017,Frugiuele2017,Flambaum2018,Berengut2018}. Precise IS measurements in atomic ions (Ca$^+$ \cite{Gebert2015,Knollmann2019}, Sr$^+$ \cite{Manovitz2019}, Ba$^+$ \cite{Imgram2019}, Ra$^+$ \cite{Holliman2019}) and neutral atoms (Sr \cite{Miyake2019}) have aimed to test King plot linearity.

The nuclei of rare-earth atoms are an interesting set that merits detailed investigation, due to their richly varied nuclear structure \cite{Angeli2013} -- among these, neodymium is an especially interesting candidate. It has five spin-zero isotopes within a convenient range of natural abundances [$^{142}$Nd (27\%), $^{144}$Nd (24\%), $^{146}$Nd (17\%), $^{148}$Nd (5.7\%) and $^{150}$Nd (5.6\%)], making it one of a handful of elements with such a set of isotopes \cite{Meija2016}. (At least four spin-zero isotopes are needed to search for new light bosons \cite{Delaunay2017}. A notable feature of this set of isotopes is a shape transition going from spherical (${}^{142}$Nd) to deformed (${}^{150}$Nd) nuclei \cite{Pitz1990}. In addition to furnishing an interesting chain of isotopes across which King plot linearity can be tested, it is plausible that the shape transition in Nd could itself be sensitive to new short-range physics. 

In this work we present our measurements of isotope shifts and King plot linearity in Nd$^+$, using a recently developed technique which enables laser absorption spectroscopy of cold ions in a neutral plasma \cite{Bhatt2019}. The IS of two transitions in Nd$^+$ were studied: $4f^4 6s \, {}^6I_{7/2}  \rightarrow  4f^4 6p \, {}^6I^{\circ}_{7/2}$ at 399 nm, and $4f^4 6s \, {}^6I_{7/2} \rightarrow [25138.550]^{\circ}_{7/2}$ at 397 nm. (We refer to these transitions by their wavelengths in the remainder of the paper.) Previously, King plot analyses of Nd$^+$ isotope shifts were reported in Refs.\ \cite{King1973, Blaise1984,Nakhate1997,Hongliang1997,Koczorowski2005,Rosner2005}, and the best IS measurements in Nd$^+$ using collinear laser ion spectroscopy obtained $\sim$50 MHz accuracy in Ref.\ \cite{Koczorowski2005}, and $\sim$3 MHz accuracy in Ref.\ \cite{Rosner2005}. Our IS measurements of the 397 nm and 399 nm transitions discussed in Section \ref{sec:measurements} improve upon Refs.\ \cite{Blaise1984,Nakhate1997} by two orders of magnitude, and are the most accurate IS measurements in Nd$^+$ to the best of our knowledge. In Section \ref{sec:clocks}, we discuss the prospects for using dipole-forbidden optical clock transitions in Nd$^+$ to test King plot linearity with further improved accuracy.

\section{Measurements}\label{sec:measurements}

\begin{figure}[b]
\includegraphics[width=\columnwidth]{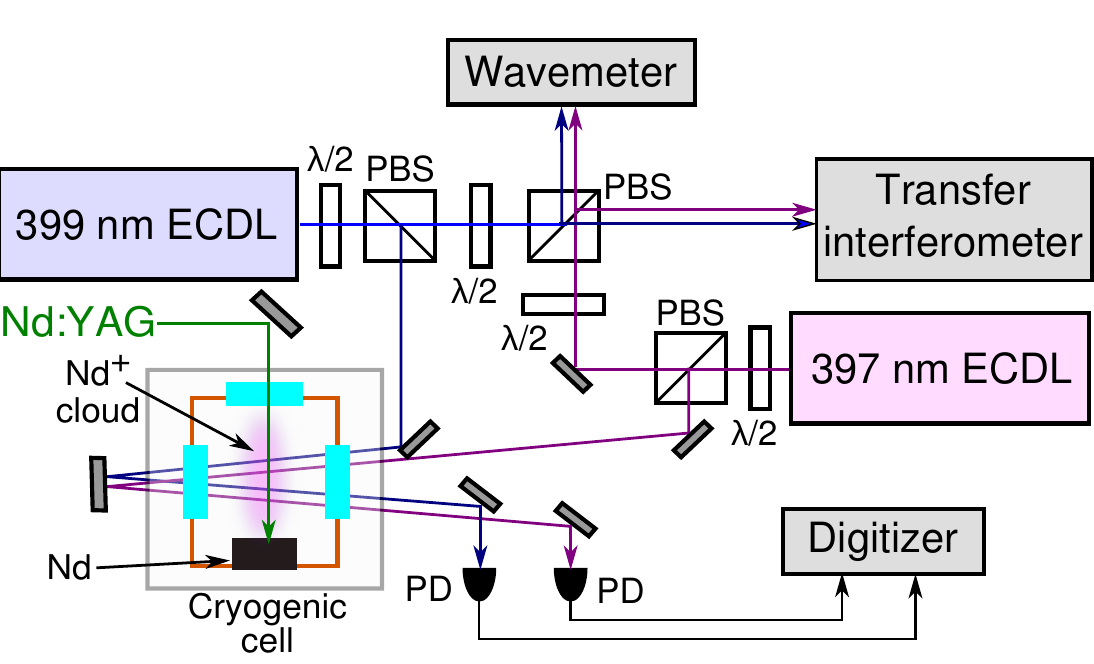} 
\caption{Schematic of the experimental setup for laser absorption spectroscopy of Nd$^+$ ions. (ECDL = external-cavity diode laser, $\lambda /2$ = half-wave plate, PBS = polarizing beam splitter, PD = photodiode.)}
\label{fig:exp_setup}
\end{figure}

The IS measurements used laser absorption spectroscopy of a cryogenic neutral plasma, produced in the apparatus described in Ref.\ \cite{Bhatt2019}. A schematic of the experimental setup is shown in Fig.\ \ref{fig:exp_setup}. External-cavity diode lasers (ECDLs) were used for spectroscopy of the 399 nm and 397 nm transitions. A portion of the light from each ECDL was sent to a wavemeter for coarse wavelength measurements, and to a transfer interferometer for frequency stabilization \cite{Jackson2018}. Nd$^+$ ion clouds were produced by pulsed Nd:YAG laser ablation of a neodymium metal target, and cooled by collisions with helium buffer gas within a cryogenically-cooled cell \cite{Bhatt2019}. Helium gas was continuously flowed into the cell (and exited from a 3 mm aperture) at a flow rate set by a mass flow controller, resulting in a nominal steady-state buffer gas density of $n_\mrm{He} \approx 10^{16} \, \mrm{cm}^{-3}$ at a temperature of 7 K. Typical optical depths measured for the Nd$^+$ clouds were between 0.1-0.5 (through a 56 mm long absorption column) depending on the energy of the ablation pulse. The laser beams for absorption spectroscopy were sent through the ion cloud, and the transmission of the cell was monitored with photodiodes and digitized for analysis. 

\begin{figure}[h]
\includegraphics[width=0.95\columnwidth]{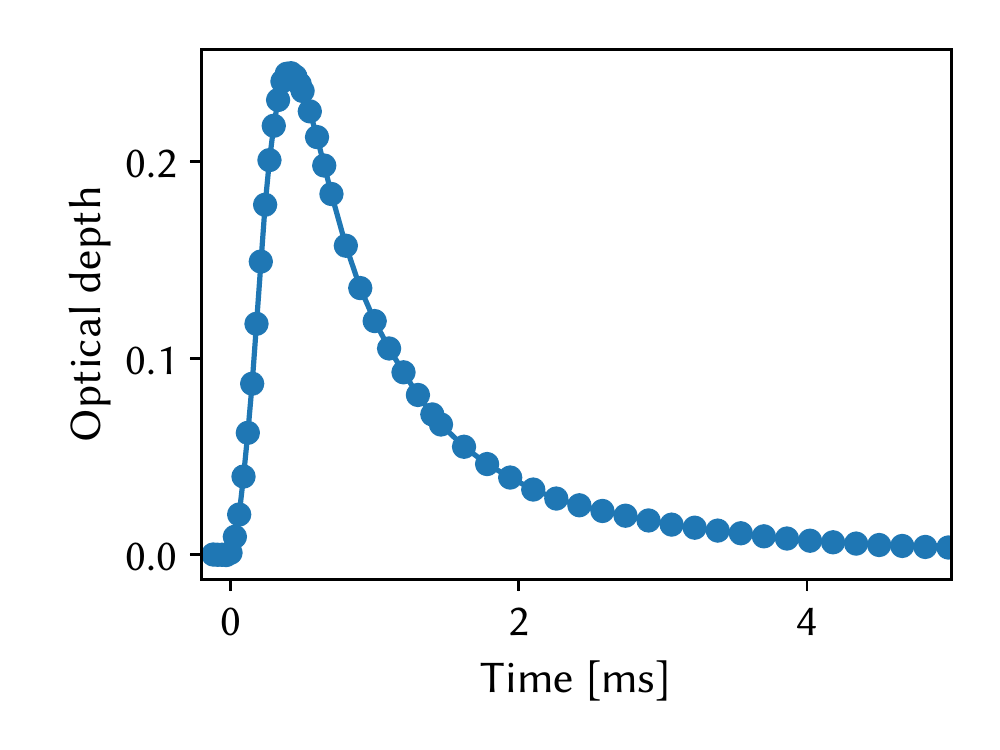}

\caption{Time evolution of the optical depth for the 397 nm transition, following the laser ablation pulse. The optical depth falls off after $\sim$ 1 ms due to recombination and diffusion of ions in the plasma. The peak value of the optical depth is used to obtain the spectra shown in Figs.\ \ref{fig:norm} and \ref{fig:combined}.}
\label{fig:absorption_time_trace}
\end{figure}

During the experiment, both lasers were simultaneously locked to the transfer interferometer (with long-term instability below 1 MHz) \cite{Jackson2018}. The frequency of one of the lasers (``probe'') was scanned across the different isotope resonances, by adjusting the phase of the interferometer signal. During this scan, the frequency of the other laser (``reference'') was fixed at the peak of a single isotope resonance. We simultaneously recorded absorption traces for each laser for 10 ms after every ablation pulse. These traces were averaged and used to obtain the peak optical depth (OD) for the probe laser. Fig.\ \ref{fig:absorption_time_trace} shows an example of the OD as a function of time, measured at the $^{142}$Nd peak on the 397 nm transition. The cell temperature, buffer gas density, and ablation laser power were recorded for each absorption trace, and used to study systematic shifts in the resonance line centers.

\begin{figure}[t]
\includegraphics[width=\columnwidth]{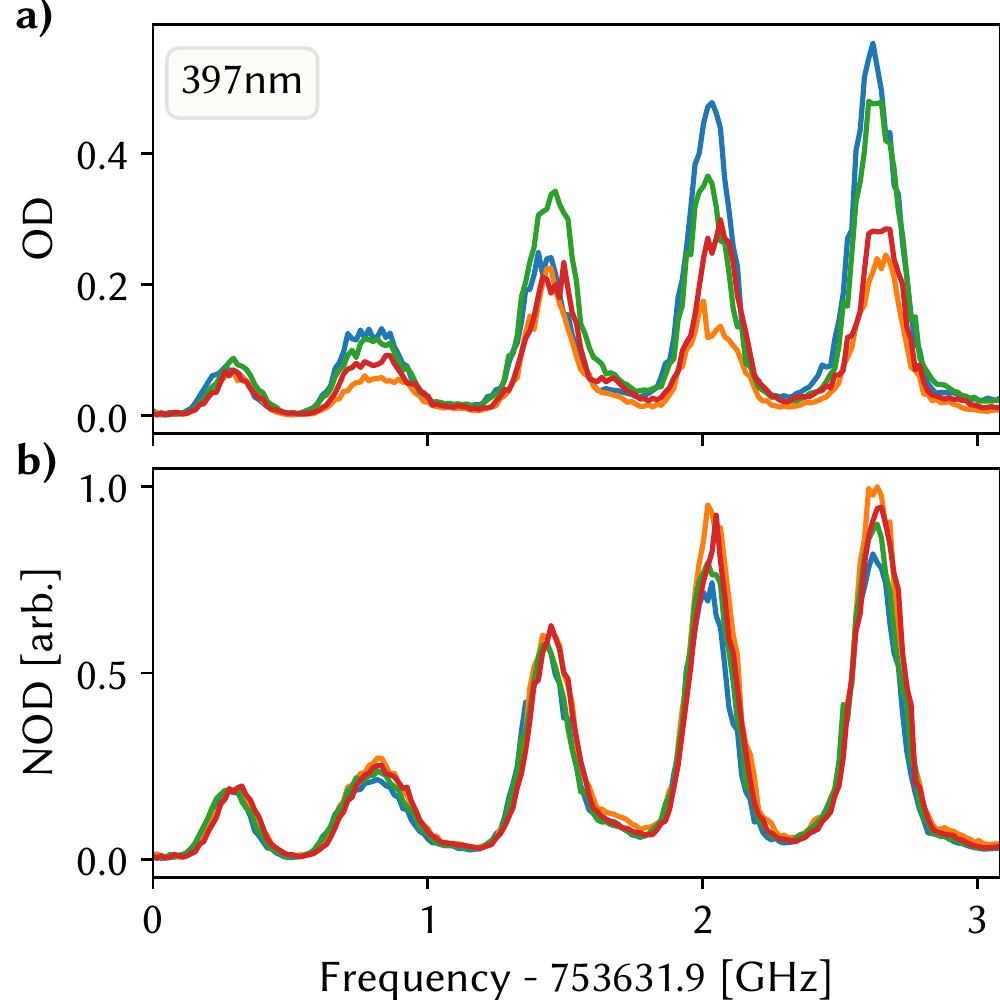}
\caption{Normalization of the optical depths. The plots in (a) show the raw optical depths measured with a 397 nm probe laser, while the plots in (b) show the probe laser optical depth normalized against the absorption of a reference laser tuned to the 399 nm transition, to reduce shot-to-shot fluctuations and drifts. The different colored traces are consecutive frequency scans over the same interval, acquired while the pulsed laser ablates a fixed spot on the metal target.}
\label{fig:norm}
\end{figure}

A source of instability in the measured OD, from one ablation pulse to the next, was the degradation of the Nd metal target surface. The resulting drift in the signal size occurred on a timescale comparable to the duration of a frequency scan over the isotope-shifted resonances, and could potentially have distorted the lineshapes. The random etching of the target surface by laser ablation also led to fluctuations from pulse to pulse, which reduced the signal to noise ratio (SNR) and made it necessary to average a large number of measurements. We suppressed both of these effects by using the reference laser signal as a measure of the ion number, and normalizing the probe laser signal to the reference. The normalized optical depth, $\mrm{NOD} = {\text{OD}_\mathrm{probe}}/{\text{OD}_\mathrm{ref}}$, had significantly better stability and SNR as shown in the example in Fig.\ \ref{fig:norm}.

\begin{figure}[h]
    \centering
    \includegraphics[width=\columnwidth]{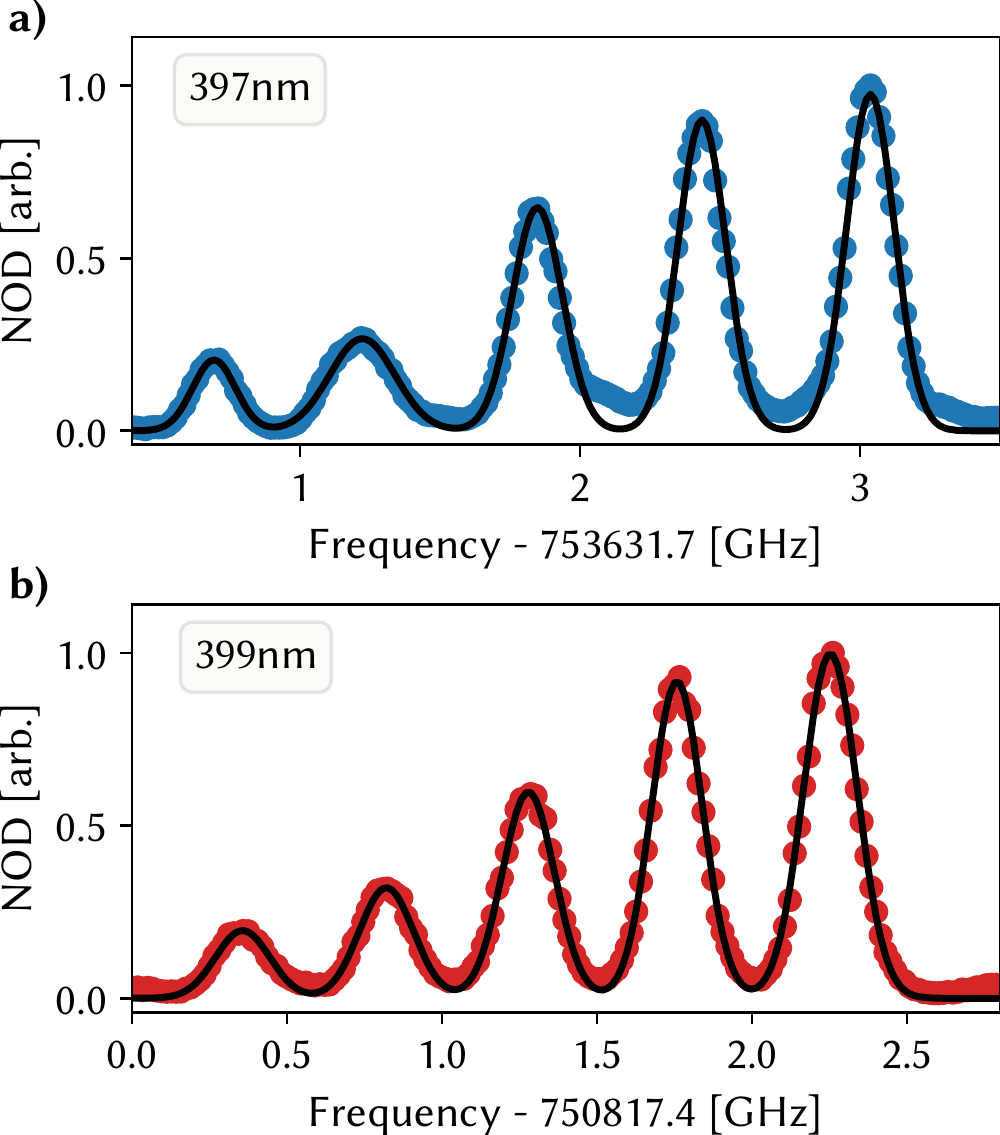}
    \caption{Isotope shift spectra of the (a) 397 nm and (b) 399 nm transitions in Nd$^+$. The markers are data points from the average of 12 frequency scans. The solid lines are a fit to a sum of five gaussians.}
    \label{fig:combined}
\end{figure}

The frequency axes in Fig.\ \ref{fig:norm} and Fig.\ \ref{fig:combined} were calibrated using the transfer interferometer and a wavemeter. Briefly, the transfer interferometer (a scanning Mach-Zehnder interferometer) was used to stabilize the laser frequency by locking the phase of the laser's interference fringe. (A detailed description of this method can be found in Ref.\ \cite{Jackson2018}.) The lock point phase could be varied digitally to scan the frequency of the laser. The lock point phase was converted to laser frequency by scanning the phase over a wide range ($\sim 5 \times 2 \pi$ rad), and measuring the corresponding frequencies on a wavemeter. The calibration was highly linear over the range of interest, and the corresponding phase-to-frequency conversion factors were determined to be $\mathcal{F}_\mrm{397} = 120.20(14)$ MHz/rad for the 397 nm laser, and $\mathcal{F}_\mrm{399} = 120.60(5)$ MHz/rad for the 399 nm laser. The uncertainty in the IS measurements due to the frequency calibration was negligible compared to the uncertainty from the residual Doppler broadening.

%\begin{table}[h]
%\begin{tabular}{|c|c|c|}
%\hline
%\multicolumn{1}{|l|}{} & 397 nm      & 399 nm      \\ \hline
%$\mathcal{F}$ {[}MHz/rad{]}      & $120.20 \pm 000.14$ & $120.60 \pm 000.05$ \\ \hline
%$f_0$ {[}GHz{]}      & $753,631.8 \pm 0.2$ & $750,815.9 \pm 0.2$ \\ \hline
%Red. Chi-Sq.           & $0.61$        & $0.18$        \\ \hline
%\end{tabular}
%\caption{Results from linear fit to obtain phase-to-frequency conversion factor $\mathcal{F}$ for 397 nm and 399 nm lasers. The uncertainty in the absolute frequency $f_0$ is limited by the wavemeter's accuracy.}\label{tab:F}
%\end{table}

Repeated measurements of the NOD versus frequency were averaged together to produce the spectra shown in Fig.\  \ref{fig:combined}. The heights of the five peaks are in good agreement with the abundances of the spin-zero Nd isotopes \cite{Meija2016}. The odd isotopes $^{143,145}\mathrm{Nd}$ ($I=7/2$) could not be discerned in our measurements as their population is distributed over a large number of hyperfine states, instead of being concentrated into a single peak as with the spin-zero isotopes \cite{Rosner2005}. 

The IS spectra were fit to a sum of five gaussians using least squares to determine the line centers. Likely due to contamination from the $^{143,145}$Nd hyperfine lines, the tails of the peaks do not completely go to zero. We accounted for this by using data from the middle of each peak (up to $\pm 1 \sigma$) to fit the data. 
%Hence the free parameters of the fit for the 397 nm line were the amplitude and line center for each peak, the width of $^{148}$Nd$^+$ and width of all other peaks. For the 399 nm line, there were no such anomalies. Consequently, the fit parameters for the 399 nm transition were: amplitudes and line centers for each peak, and a single width for all peaks. 

From the widths of the gaussian fits, the inferred temperature of the ion clouds was around 18 K. Lower ablation pulse energies led to better equilibration between the ion cloud and the cryogenic cell (as demonstrated in \cite{Bhatt2019}), but yielded reduced signals with higher fluctuations. Therefore the measurements reported here used higher pulse energies to obtain better SNR. The difference in the line centers of isotopes with mass $m_A$ and $m_{A'}$ was used to compute the isotope shifts $\delta\nu_{\mathrm{397nm}}^{AA'}$ and $\delta\nu_{\mathrm{399nm}}^{AA'}$ listed in Table \ref{tab:sys}. Uncertainties listed for the isotope shifts are derived from the fit uncertainties to the resonance peaks. 

To study systematic errors, we varied the ablation laser pulse energy, the buffer gas density (by varying the helium gas flow rate through the cell), and the probe laser power. The Nd:YAG pulse energy was varied between 15 mJ and 21 mJ, the buffer gas density between $4\times10^{15}$ cm$^{-3}$ and $2\times10^{16}$ cm$^{-3}$, and the probe laser power between 100 $\mu$W and 400 $\mu$W. Higher energy ablation pulses resulted in higher ion cloud temperatures, but did not measurably affect the isotope shifts. The buffer gas density and the probe laser power did not affect them either. These tests, summarized in Table \ref{tab:sys}, lead us to conclude that systematic errors due to, e.g., collisional shifts, Doppler lineshape corrections and optical pumping effects are significantly smaller than the statistical uncertainties.
\setlength{\tabcolsep}{6pt}
\begin{table*}[htbp]
\centering
\begin{tabularx}{\textwidth}{@{} cccXXXXX @{}}
\toprule
Ablation pulse energy & He flow rate & Probe laser power& 
\multicolumn{5}{c}{Pairwise isotope shift $\delta\nu_{\lambda}^{AA'}$ [GHz]} \\ 
$[$mJ$]$& [sccm]& [$\mu$W]& 142-146  & 142-150  & 142-148  & 144-148  & 150-148 \\ 
\midrule
\multicolumn{8}{l}{397 nm} \\ \midrule
$17.0$  & $0.70$  & $100$  & $1.191 (5)$  & $2.345 (4)$  & $1.819 (5)$  & $1.216 (4)$  & $-0.526 (3)$\\
$21.0$  & $0.70$  & $100$  & $1.193 (4)$  & $2.344 (4)$  & $1.817 (4)$  & $1.214 (3)$  & $-0.527 (2)$\\
$21.0$  & $0.70$  & $400$  & $1.191 (4)$  & $2.344 (3)$  & $1.819 (4)$  & $1.214 (3)$  & $-0.526 (2)$\\
$17.0$  & $0.35$  & $100$  & $1.182 (4)$  & $2.332 (4)$  & $1.803 (4)$  & $1.204 (3)$  & $-0.530 (2)$\\
$17.0$  & $1.40$  & $100$  & $1.196 (5)$  & $2.344 (4)$  & $1.820 (4)$  & $1.202 (4)$  & $-0.524 (2)$\\
$15.4$  & $0.70$  & $100$  & $1.193 (5)$  & $2.339 (4)$  & $1.816 (5)$  & $1.211 (4)$  & $-0.524 (3)$\\
$18.6$  & $0.70$  & $100$  & $1.193 (5)$  & $2.342 (4)$  & $1.820 (4)$  & $1.214 (4)$  & $-0.522 (3)$\\ \midrule
\multicolumn{8}{l}{399 nm} \\ \midrule
$17.0$  & $0.70$  & $100$  & $0.978 (2)$  & $1.896 (2)$  & $1.433 (2)$  & $0.935 (2)$  & $-0.463 (2)$\\
$21.0$  & $0.70$  & $100$  & $0.973 (3)$  & $1.890 (3)$  & $1.427 (3)$  & $0.934 (2)$  & $-0.463 (3)$\\
$21.0$  & $0.70$  & $400$  & $0.980 (2)$  & $1.896 (2)$  & $1.433 (2)$  & $0.937 (2)$  & $-0.462 (1)$\\
$15.4$  & $0.70$  & $100$  & $0.979 (2)$  & $1.894 (2)$  & $1.427 (2)$  & $0.934 (2)$  & $-0.467 (2)$\\
$17.0$  & $0.35$  & $100$  & $0.974 (2)$  & $1.891 (2)$  & $1.429 (2)$  & $0.939 (1)$  & $-0.462 (1)$\\
$17.0$  & $1.40$  & $100$  & $0.972 (2)$  & $1.897 (2)$  & $1.431 (2)$  & $0.937 (2)$  & $-0.466 (2)$\\
$18.6$  & $0.70$  & $100$  & $0.976 (2)$  & $1.898 (1)$  & $1.432 (1)$  & $0    .937 (1)$  & $-0.467 (1)$\\ \bottomrule
\end{tabularx}
\caption{Isotope shifts of the 397 nm and 399 nm transitions for various values of the Nd:YAG ablation pulse energy, helium buffer gas flow rate (in units of standard cc per minute), and probe laser power.}
\label{tab:sys}
\end{table*}

\begin{figure}[htbp]
    \centering
    \includegraphics[width=\columnwidth]{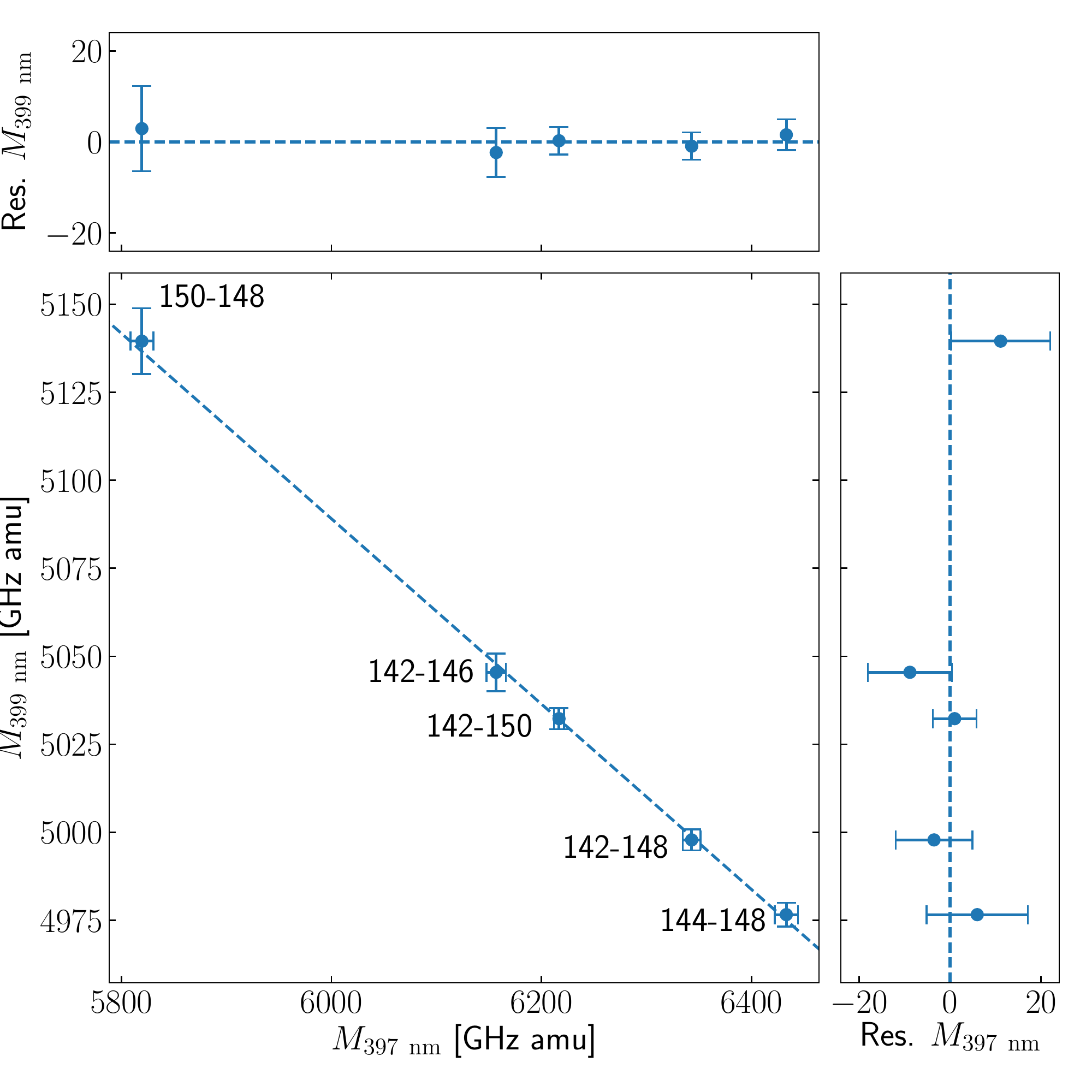}
    \caption{King plot of the modified isotope shifts of the 397 nm and 399 nm transitions. The dashed straight line fits with a slope of $-0.263\pm0.006$ and y-intercept $6669\pm36$ GHz amu. The top and side panels show the fit residuals. The plot is linear to within the experimental resolution.}
%The reduced-$\chi^2$ of the fit was 0.2 for $m\delta\nu^{c}_{399 nm}$ and 0.8 for $m\delta\nu^{c}_{397 nm}$.}
    \label{fig:kingplot}
\end{figure}

The average isotope shifts $\delta\nu_{\lambda}^{AA'}$ from Table \ref{tab:sys}, for each pair of isotopes $A, A'$ and wavelength $\lambda$, were used to compute the modified isotope shifts (MIS), $M_{\lambda}^{AA'} = \delta\nu_{\lambda}^{AA'} \, g^{AA'}$, where $g^{AA'}=(1/m_A-1/m_{A'})^{-1}$ is a factor that depends on the masses $m_A, m_A'$ of the pair of isotopes. The atomic masses of the neodymium isotopes from Ref.\ \cite{Wang2012} were used in the calculation of $g^{AA'}$. The MIS for the 397 nm and 399 nm transitions are plotted against each other in the King plot shown in Fig.\ \ref{fig:kingplot}. As is evident from the residuals to the straight line fit, the King plot has no detectable nonlinearity.

% \begin{table}[]
% \begin{tabular}{|c|c|c|}
% \hline
% Isotope-Shift $\nu_{AA'}$ & 397nm [GHz]      & 399nm [GHz]  \\ \hline
% 142-146    & $1.1910 (18)$    & $0.9760 (10)$    \\ \hline
% 142-150    & $2.3415 (18)$    & $1.8954 (11)$    \\ \hline
% 142-148    & $1.8159 (24)$    & $1.4308 (9)$    \\ \hline
% 144-148    & $1.2109 (21)$    & $0.9367 (6)$    \\ \hline
% 150-148    & $-0.5259 (10)$    & $-0.4645 (8)$    \\ \hline
% \end{tabular}
% \caption{Isotope shifts of Nd$^+$ for the 399nm and the 397nm transitions calculated by fitting a constant for each shift across a range of experiment parameters.}\label{tab:IS}
% \end{table}

\section{Prospects for high-precision isotope shift measurements} \label{sec:clocks} %\element{Nd}$^+$}
The precision of the IS measurements described in the previous section is primarily limited by the Doppler broadening due to the residual thermal motion of ions in the cryogenic cell. However, even with a laser-cooled or trapped sample of ions, the transitions studied here are not likely to lead to significant improvements as they are $E1$-allowed transitions with relatively large linewidths.

Sizeable improvements to the precision of IS measurements in Nd$^+$ can be made using spectroscopy of narrow-linewidth transitions in an ion trap. (Yb$^+$ for example, with its comparable mass, can be used to sympathetically cool a co-trapped Nd$^+$ ion.) Nd$^+$ has a set of four narrow transitions, which could be used to improve the precision of King plot linearity tests. Uniquely, these transitions lie in the telecommunications C-band (1530-1560$\,$nm) when they are driven as degenerate two-photon transitions. (These transitions could therefore also be useful for constructing optical clocks, whose signals can be directly transmitted over telecom networks without an intermediary frequency comb.) Spectroscopy of C-band clock transitions is technologically convenient due to the availability of high-quality commercial optical components, and is well-suited to take advantage of recent advances in high-finesse optical cavity technology \cite{Matei2017,Robinson2019}.

Table \ref{tab:clock_transitions} lists four forbidden optical transitions that could lead to high-precision IS measurements. Two of the transitions are inner shell intra-configuration $f^4 \to f^4$ transitions, whereas the other two are outer electron $s \to d$ transitions. This is an advantageous fact: comparisons between transitions of different types will enable good discrimination between the field-shift and mass-shift contributions to the IS; on the other hand, comparisons between the pairs of similar transitions will allow for careful tests of systematics. Spectroscopy of this set of clock transitions thus offers the prospect of IS measurements in Nd$^+$ with sub-hertz precision and accuracy, with corresponding improvements to the physics reach of King plot linearity tests.

\setlength{\tabcolsep}{6pt}
\begin{table*}[htbp]
    \centering
    \begin{tabularx}{\textwidth}{@{} lllll @{}}
    \toprule
          Excited state & $J_e$ & Energy [cm$^{-1}$] & 2$\gamma$ wavelength [nm] & Notes \\ \midrule
          $4f^4 6s$ & $3/2$ &12747.61& 1568.92 & $f^4\rightarrow f^4$ intra-configuration transition, $\Delta J=2$ \\
        %   $4f^35d6s$ & $7/2$ & 12861.42 & 1555.04 &  $f\rightarrow d$ transition, $E1$ allowed \\
          $4f^4 6s$ & $7/2$ & 12879.05 &1552.91 & $f^4\rightarrow f^4$ intra-configuration transition, $\Delta J=0$ \\
          $4f^4 5d$ & $11/2$ & 12887.09 & 1551.94 & $s\rightarrow d$ transition, $\Delta J=2$\\
          $4f^4 5d$ & $15/2$ & 12906.57 & 1549.60 & $s\rightarrow d$ transition, $\Delta J=4$ \\
          \bottomrule
    \end{tabularx}
    \caption{Excited states for a set of forbidden transitions in Nd$^+$, whose two-photon excitation wavelengths lie within the telecommunications C-band. The excited state angular momentum quantum number is $J_e$ and the energy is relative to the $4f^4 6s\,^6I_{7/2}$ ground state. Data on the energies are from the NIST Atomic Spectra Database \cite{NIST_ASD}.}
    \label{tab:clock_transitions}
\end{table*}

\section{Summary}
We have measured the isotope shifts of the $4f^46s\rightarrow [25044.7]^{\circ}_{7/2}$ and $4f^46s\rightarrow [25138.6]^{\circ}_{7/2}$ transitions in Nd$^+$ with significantly improved accuracy. Laser absorption measurements with good SNR were enabled by the high density of cold ions within a cryogenically-cooled neutral plasma. Our measurements test King plot linearity over a set of spin-zero isotopes that spans the nuclear shape transition in neodymium, and they could be useful for comparisons against \textit{ab initio} calculations for rare-earth ions, and in searches for new physics. We have also identified a set of optical clock transitions that could be used for high-precision isotope shift measurements in Nd$^+$.

~ \emph{Acknowledgments. --} We thank Victor Flambaum and Andrew Jayich for helpful discussions. We acknowledge the contributions of Anthony Roitman and Monica Zhu. This research was supported by NSERC, the Canada Foundation for Innovation and the Ontario Research Fund. ACV acknowledges support from Canada Research Chairs and a Sloan Research Fellowship.

\bibliography{IsotopeShifts_NdII.bib}

\begin{thebibliography}{35}%
\makeatletter
\providecommand \@ifxundefined [1]{%
 \@ifx{#1\undefined}
}%
\providecommand \@ifnum [1]{%
 \ifnum #1\expandafter \@firstoftwo
 \else \expandafter \@secondoftwo
 \fi
}%
\providecommand \@ifx [1]{%
 \ifx #1\expandafter \@firstoftwo
 \else \expandafter \@secondoftwo
 \fi
}%
\providecommand \natexlab [1]{#1}%
\providecommand \enquote  [1]{``#1''}%
\providecommand \bibnamefont  [1]{#1}%
\providecommand \bibfnamefont [1]{#1}%
\providecommand \citenamefont [1]{#1}%
\providecommand \href@noop [0]{\@secondoftwo}%
\providecommand \href [0]{\begingroup \@sanitize@url \@href}%
\providecommand \@href[1]{\@@startlink{#1}\@@href}%
\providecommand \@@href[1]{\endgroup#1\@@endlink}%
\providecommand \@sanitize@url [0]{\catcode `\\12\catcode `\$12\catcode
  `\&12\catcode `\#12\catcode `\^12\catcode `\_12\catcode `\%12\relax}%
\providecommand \@@startlink[1]{}%
\providecommand \@@endlink[0]{}%
\providecommand \url  [0]{\begingroup\@sanitize@url \@url }%
\providecommand \@url [1]{\endgroup\@href {#1}{\urlprefix }}%
\providecommand \urlprefix  [0]{URL }%
\providecommand \Eprint [0]{\href }%
\providecommand \doibase [0]{http://dx.doi.org/}%
\providecommand \selectlanguage [0]{\@gobble}%
\providecommand \bibinfo  [0]{\@secondoftwo}%
\providecommand \bibfield  [0]{\@secondoftwo}%
\providecommand \translation [1]{[#1]}%
\providecommand \BibitemOpen [0]{}%
\providecommand \bibitemStop [0]{}%
\providecommand \bibitemNoStop [0]{.\EOS\space}%
\providecommand \EOS [0]{\spacefactor3000\relax}%
\providecommand \BibitemShut  [1]{\csname bibitem#1\endcsname}%
\let\auto@bib@innerbib\@empty
%</preamble>
\bibitem [{\citenamefont {Angeli}\ and\ \citenamefont
  {Marinova}(2013)}]{Angeli2013}%
  \BibitemOpen
  \bibfield  {author} {\bibinfo {author} {\bibfnamefont {I.}~\bibnamefont
  {Angeli}}\ and\ \bibinfo {author} {\bibfnamefont {K.}~\bibnamefont
  {Marinova}},\ }\href {\doibase https://doi.org/10.1016/j.adt.2011.12.006}
  {\bibfield  {journal} {\bibinfo  {journal} {Atomic Data and Nuclear Data
  Tables}\ }\textbf {\bibinfo {volume} {99}},\ \bibinfo {pages} {69 } (\bibinfo
  {year} {2013})}\BibitemShut {NoStop}%
\bibitem [{\citenamefont {Blaum}\ \emph {et~al.}(2013)\citenamefont {Blaum},
  \citenamefont {Dilling},\ and\ \citenamefont
  {N{\"{o}}rtersh{\"{a}}user}}]{Blaum2013}%
  \BibitemOpen
  \bibfield  {author} {\bibinfo {author} {\bibfnamefont {K.}~\bibnamefont
  {Blaum}}, \bibinfo {author} {\bibfnamefont {J.}~\bibnamefont {Dilling}}, \
  and\ \bibinfo {author} {\bibfnamefont {W.}~\bibnamefont
  {N{\"{o}}rtersh{\"{a}}user}},\ }\href {\doibase
  10.1088/0031-8949/2013/t152/014017} {\bibfield  {journal} {\bibinfo
  {journal} {Phys. Scripta}\ }\textbf {\bibinfo {volume} {T152}},\ \bibinfo
  {pages} {014017} (\bibinfo {year} {2013})}\BibitemShut {NoStop}%
\bibitem [{\citenamefont {Campbell}\ \emph {et~al.}(2016)\citenamefont
  {Campbell}, \citenamefont {Moore},\ and\ \citenamefont
  {Pearson}}]{Campbell2016}%
  \BibitemOpen
  \bibfield  {author} {\bibinfo {author} {\bibfnamefont {P.}~\bibnamefont
  {Campbell}}, \bibinfo {author} {\bibfnamefont {I.}~\bibnamefont {Moore}}, \
  and\ \bibinfo {author} {\bibfnamefont {M.}~\bibnamefont {Pearson}},\ }\href
  {\doibase https://doi.org/10.1016/j.ppnp.2015.09.003} {\bibfield  {journal}
  {\bibinfo  {journal} {Prog. Part. Nucl. Phys.}\ }\textbf {\bibinfo {volume}
  {86}},\ \bibinfo {pages} {127 } (\bibinfo {year} {2016})}\BibitemShut
  {NoStop}%
\bibitem [{\citenamefont {Kalita}\ \emph {et~al.}(2018)\citenamefont {Kalita}
  \emph {et~al.}}]{Kalita2018}%
  \BibitemOpen
  \bibfield  {author} {\bibinfo {author} {\bibfnamefont {M.~R.}\ \bibnamefont
  {Kalita}} \emph {et~al.},\ }\href {\doibase 10.1103/PhysRevA.97.042507}
  {\bibfield  {journal} {\bibinfo  {journal} {Phys. Rev. A}\ }\textbf {\bibinfo
  {volume} {97}},\ \bibinfo {pages} {042507} (\bibinfo {year}
  {2018})}\BibitemShut {NoStop}%
\bibitem [{\citenamefont {Ohayon}\ \emph {et~al.}(2019)\citenamefont {Ohayon},
  \citenamefont {Rahangdale}, \citenamefont {Geddes}, \citenamefont
  {Berengut},\ and\ \citenamefont {Ron}}]{Ohayon2019}%
  \BibitemOpen
  \bibfield  {author} {\bibinfo {author} {\bibfnamefont {B.}~\bibnamefont
  {Ohayon}}, \bibinfo {author} {\bibfnamefont {H.}~\bibnamefont {Rahangdale}},
  \bibinfo {author} {\bibfnamefont {A.~J.}\ \bibnamefont {Geddes}}, \bibinfo
  {author} {\bibfnamefont {J.~C.}\ \bibnamefont {Berengut}}, \ and\ \bibinfo
  {author} {\bibfnamefont {G.}~\bibnamefont {Ron}},\ }\href@noop {} {\bibfield
  {journal} {\bibinfo  {journal} {Phys. Rev. A}\ }\textbf {\bibinfo {volume}
  {99}} (\bibinfo {year} {2019})}\BibitemShut {NoStop}%
\bibitem [{\citenamefont {Dzuba}\ \emph {et~al.}(2017)\citenamefont {Dzuba},
  \citenamefont {Flambaum},\ and\ \citenamefont {Webb}}]{Dzuba2017b}%
  \BibitemOpen
  \bibfield  {author} {\bibinfo {author} {\bibfnamefont {V.~A.}\ \bibnamefont
  {Dzuba}}, \bibinfo {author} {\bibfnamefont {V.~V.}\ \bibnamefont {Flambaum}},
  \ and\ \bibinfo {author} {\bibfnamefont {J.~K.}\ \bibnamefont {Webb}},\
  }\href {\doibase 10.1103/PhysRevA.95.062515} {\bibfield  {journal} {\bibinfo
  {journal} {Phys. Rev. A}\ }\textbf {\bibinfo {volume} {95}},\ \bibinfo
  {pages} {062515} (\bibinfo {year} {2017})}\BibitemShut {NoStop}%
\bibitem [{\citenamefont {Kozlov}\ \emph {et~al.}(2004)\citenamefont {Kozlov},
  \citenamefont {Korol}, \citenamefont {Berengut}, \citenamefont {Dzuba},\ and\
  \citenamefont {Flambaum}}]{Kozlov2004}%
  \BibitemOpen
  \bibfield  {author} {\bibinfo {author} {\bibfnamefont {M.~G.}\ \bibnamefont
  {Kozlov}}, \bibinfo {author} {\bibfnamefont {V.~A.}\ \bibnamefont {Korol}},
  \bibinfo {author} {\bibfnamefont {J.~C.}\ \bibnamefont {Berengut}}, \bibinfo
  {author} {\bibfnamefont {V.~A.}\ \bibnamefont {Dzuba}}, \ and\ \bibinfo
  {author} {\bibfnamefont {V.~V.}\ \bibnamefont {Flambaum}},\ }\href {\doibase
  10.1103/PhysRevA.70.062108} {\bibfield  {journal} {\bibinfo  {journal} {Phys.
  Rev. A}\ }\textbf {\bibinfo {volume} {70}},\ \bibinfo {pages} {062108}
  (\bibinfo {year} {2004})}\BibitemShut {NoStop}%
\bibitem [{\citenamefont {Murphy}\ and\ \citenamefont
  {Berengut}(2013)}]{Murphy2013}%
  \BibitemOpen
  \bibfield  {author} {\bibinfo {author} {\bibfnamefont {M.~T.}\ \bibnamefont
  {Murphy}}\ and\ \bibinfo {author} {\bibfnamefont {J.~C.}\ \bibnamefont
  {Berengut}},\ }\href {\doibase 10.1093/mnras/stt2204} {\bibfield  {journal}
  {\bibinfo  {journal} {Mon. Not. R. Astron. Soc.}\ }\textbf {\bibinfo {volume}
  {438}},\ \bibinfo {pages} {388} (\bibinfo {year} {2013})}\BibitemShut
  {NoStop}%
\bibitem [{\citenamefont {Delaunay}\ \emph {et~al.}(2017)\citenamefont
  {Delaunay}, \citenamefont {Ozeri}, \citenamefont {Perez},\ and\ \citenamefont
  {Soreq}}]{Delaunay2017}%
  \BibitemOpen
  \bibfield  {author} {\bibinfo {author} {\bibfnamefont {C.}~\bibnamefont
  {Delaunay}}, \bibinfo {author} {\bibfnamefont {R.}~\bibnamefont {Ozeri}},
  \bibinfo {author} {\bibfnamefont {G.}~\bibnamefont {Perez}}, \ and\ \bibinfo
  {author} {\bibfnamefont {Y.}~\bibnamefont {Soreq}},\ }\href {\doibase
  10.1103/PhysRevD.96.093001} {\bibfield  {journal} {\bibinfo  {journal} {Phys.
  Rev. D}\ }\textbf {\bibinfo {volume} {96}},\ \bibinfo {pages} {1} (\bibinfo
  {year} {2017})}\BibitemShut {NoStop}%
\bibitem [{\citenamefont {Frugiuele}\ \emph {et~al.}(2017)\citenamefont
  {Frugiuele}, \citenamefont {Fuchs}, \citenamefont {Perez},\ and\
  \citenamefont {Schlaffer}}]{Frugiuele2017}%
  \BibitemOpen
  \bibfield  {author} {\bibinfo {author} {\bibfnamefont {C.}~\bibnamefont
  {Frugiuele}}, \bibinfo {author} {\bibfnamefont {E.}~\bibnamefont {Fuchs}},
  \bibinfo {author} {\bibfnamefont {G.}~\bibnamefont {Perez}}, \ and\ \bibinfo
  {author} {\bibfnamefont {M.}~\bibnamefont {Schlaffer}},\ }\href {\doibase
  10.1103/PhysRevD.96.015011} {\bibfield  {journal} {\bibinfo  {journal} {Phys.
  Rev. D}\ }\textbf {\bibinfo {volume} {96}},\ \bibinfo {pages} {015011}
  (\bibinfo {year} {2017})}\BibitemShut {NoStop}%
\bibitem [{\citenamefont {Flambaum}\ \emph {et~al.}(2018)\citenamefont
  {Flambaum}, \citenamefont {Geddes},\ and\ \citenamefont
  {Viatkina}}]{Flambaum2018}%
  \BibitemOpen
  \bibfield  {author} {\bibinfo {author} {\bibfnamefont {V.~V.}\ \bibnamefont
  {Flambaum}}, \bibinfo {author} {\bibfnamefont {A.~J.}\ \bibnamefont
  {Geddes}}, \ and\ \bibinfo {author} {\bibfnamefont {A.~V.}\ \bibnamefont
  {Viatkina}},\ }\href {\doibase 10.1103/PhysRevA.97.032510} {\bibfield
  {journal} {\bibinfo  {journal} {Phys. Rev. A}\ }\textbf {\bibinfo {volume}
  {97}},\ \bibinfo {pages} {1} (\bibinfo {year} {2018})}\BibitemShut {NoStop}%
\bibitem [{\citenamefont {Berengut}\ \emph {et~al.}(2018)\citenamefont
  {Berengut} \emph {et~al.}}]{Berengut2018}%
  \BibitemOpen
  \bibfield  {author} {\bibinfo {author} {\bibfnamefont {J.~C.}\ \bibnamefont
  {Berengut}} \emph {et~al.},\ }\href {\doibase 10.1103/PhysRevLett.120.091801}
  {\bibfield  {journal} {\bibinfo  {journal} {Phys. Rev. Lett.}\ }\textbf
  {\bibinfo {volume} {120}},\ \bibinfo {pages} {091801} (\bibinfo {year}
  {2018})}\BibitemShut {NoStop}%
\bibitem [{\citenamefont {King}(1984)}]{King1984}%
  \BibitemOpen
  \bibfield  {author} {\bibinfo {author} {\bibfnamefont {W.~H.}\ \bibnamefont
  {King}},\ }\href@noop {} {\emph {\bibinfo {title} {Isotope Shifts in Atomic
  Spectra}}}\ (\bibinfo  {publisher} {Springer},\ \bibinfo {year}
  {1984})\BibitemShut {NoStop}%
\bibitem [{\citenamefont {Blundell}\ \emph {et~al.}(1987)\citenamefont
  {Blundell}, \citenamefont {Baird}, \citenamefont {Palmer}, \citenamefont
  {Stacey},\ and\ \citenamefont {Woodgate}}]{Blundell1987}%
  \BibitemOpen
  \bibfield  {author} {\bibinfo {author} {\bibfnamefont {S.~A.}\ \bibnamefont
  {Blundell}}, \bibinfo {author} {\bibfnamefont {P.~E.~G.}\ \bibnamefont
  {Baird}}, \bibinfo {author} {\bibfnamefont {C.~W.~P.}\ \bibnamefont
  {Palmer}}, \bibinfo {author} {\bibfnamefont {D.~N.}\ \bibnamefont {Stacey}},
  \ and\ \bibinfo {author} {\bibfnamefont {G.~K.}\ \bibnamefont {Woodgate}},\
  }\href {\doibase 10.1088/0022-3700/20/15/015} {\bibfield  {journal} {\bibinfo
   {journal} {J. Phys. B}\ }\textbf {\bibinfo {volume} {20}},\ \bibinfo {pages}
  {3663} (\bibinfo {year} {1987})}\BibitemShut {NoStop}%
\bibitem [{\citenamefont {Yerokhin}\ \emph {et~al.}(2020)\citenamefont
  {Yerokhin}, \citenamefont {M\"uller}, \citenamefont {Surzhykov},
  \citenamefont {Micke},\ and\ \citenamefont {Schmidt}}]{Yerokhin2020}%
  \BibitemOpen
  \bibfield  {author} {\bibinfo {author} {\bibfnamefont {V.~A.}\ \bibnamefont
  {Yerokhin}}, \bibinfo {author} {\bibfnamefont {R.~A.}\ \bibnamefont
  {M\"uller}}, \bibinfo {author} {\bibfnamefont {A.}~\bibnamefont {Surzhykov}},
  \bibinfo {author} {\bibfnamefont {P.}~\bibnamefont {Micke}}, \ and\ \bibinfo
  {author} {\bibfnamefont {P.~O.}\ \bibnamefont {Schmidt}},\ }\href {\doibase
  10.1103/PhysRevA.101.012502} {\bibfield  {journal} {\bibinfo  {journal}
  {Phys. Rev. A}\ }\textbf {\bibinfo {volume} {101}},\ \bibinfo {pages}
  {012502} (\bibinfo {year} {2020})}\BibitemShut {NoStop}%
\bibitem [{\citenamefont {Gebert}\ \emph {et~al.}(2015)\citenamefont {Gebert},
  \citenamefont {Wan}, \citenamefont {Wolf}, \citenamefont {Angstmann},
  \citenamefont {Berengut},\ and\ \citenamefont {Schmidt}}]{Gebert2015}%
  \BibitemOpen
  \bibfield  {author} {\bibinfo {author} {\bibfnamefont {F.}~\bibnamefont
  {Gebert}}, \bibinfo {author} {\bibfnamefont {Y.}~\bibnamefont {Wan}},
  \bibinfo {author} {\bibfnamefont {F.}~\bibnamefont {Wolf}}, \bibinfo {author}
  {\bibfnamefont {C.~N.}\ \bibnamefont {Angstmann}}, \bibinfo {author}
  {\bibfnamefont {J.~C.}\ \bibnamefont {Berengut}}, \ and\ \bibinfo {author}
  {\bibfnamefont {P.~O.}\ \bibnamefont {Schmidt}},\ }\href {\doibase
  10.1103/PhysRevLett.115.053003} {\bibfield  {journal} {\bibinfo  {journal}
  {Phys. Rev. Lett.}\ }\textbf {\bibinfo {volume} {115}},\ \bibinfo {pages}
  {053003} (\bibinfo {year} {2015})}\BibitemShut {NoStop}%
\bibitem [{\citenamefont {Knollmann}\ \emph {et~al.}(2019)\citenamefont
  {Knollmann}, \citenamefont {Patel},\ and\ \citenamefont
  {Doret}}]{Knollmann2019}%
  \BibitemOpen
  \bibfield  {author} {\bibinfo {author} {\bibfnamefont {F.~W.}\ \bibnamefont
  {Knollmann}}, \bibinfo {author} {\bibfnamefont {A.~N.}\ \bibnamefont
  {Patel}}, \ and\ \bibinfo {author} {\bibfnamefont {S.~C.}\ \bibnamefont
  {Doret}},\ }\href {\doibase 10.1103/PhysRevA.100.022514} {\bibfield
  {journal} {\bibinfo  {journal} {Phys. Rev. A}\ }\textbf {\bibinfo {volume}
  {100}},\ \bibinfo {pages} {022514} (\bibinfo {year} {2019})}\BibitemShut
  {NoStop}%
\bibitem [{\citenamefont {Manovitz}\ \emph {et~al.}(2019)\citenamefont
  {Manovitz}, \citenamefont {Shaniv}, \citenamefont {Shapira}, \citenamefont
  {Ozeri},\ and\ \citenamefont {Akerman}}]{Manovitz2019}%
  \BibitemOpen
  \bibfield  {author} {\bibinfo {author} {\bibfnamefont {T.}~\bibnamefont
  {Manovitz}}, \bibinfo {author} {\bibfnamefont {R.}~\bibnamefont {Shaniv}},
  \bibinfo {author} {\bibfnamefont {Y.}~\bibnamefont {Shapira}}, \bibinfo
  {author} {\bibfnamefont {R.}~\bibnamefont {Ozeri}}, \ and\ \bibinfo {author}
  {\bibfnamefont {N.}~\bibnamefont {Akerman}},\ }\href {\doibase
  10.1103/PhysRevLett.123.203001} {\bibfield  {journal} {\bibinfo  {journal}
  {Phys. Rev. Lett.}\ }\textbf {\bibinfo {volume} {123}},\ \bibinfo {pages}
  {203001} (\bibinfo {year} {2019})}\BibitemShut {NoStop}%
\bibitem [{\citenamefont {Imgram}\ \emph {et~al.}(2019)\citenamefont {Imgram},
  \citenamefont {K\"onig}, \citenamefont {Kr\"amer}, \citenamefont {Ratajczyk},
  \citenamefont {M\"uller}, \citenamefont {Surzhykov},\ and\ \citenamefont
  {N\"ortersh\"auser}}]{Imgram2019}%
  \BibitemOpen
  \bibfield  {author} {\bibinfo {author} {\bibfnamefont {P.}~\bibnamefont
  {Imgram}}, \bibinfo {author} {\bibfnamefont {K.}~\bibnamefont {K\"onig}},
  \bibinfo {author} {\bibfnamefont {J.}~\bibnamefont {Kr\"amer}}, \bibinfo
  {author} {\bibfnamefont {T.}~\bibnamefont {Ratajczyk}}, \bibinfo {author}
  {\bibfnamefont {R.~A.}\ \bibnamefont {M\"uller}}, \bibinfo {author}
  {\bibfnamefont {A.}~\bibnamefont {Surzhykov}}, \ and\ \bibinfo {author}
  {\bibfnamefont {W.}~\bibnamefont {N\"ortersh\"auser}},\ }\href {\doibase
  10.1103/PhysRevA.99.012511} {\bibfield  {journal} {\bibinfo  {journal} {Phys.
  Rev. A}\ }\textbf {\bibinfo {volume} {99}},\ \bibinfo {pages} {012511}
  (\bibinfo {year} {2019})}\BibitemShut {NoStop}%
\bibitem [{\citenamefont {Holliman}\ \emph {et~al.}(2019)\citenamefont
  {Holliman}, \citenamefont {Fan},\ and\ \citenamefont
  {Jayich}}]{Holliman2019}%
  \BibitemOpen
  \bibfield  {author} {\bibinfo {author} {\bibfnamefont {C.~A.}\ \bibnamefont
  {Holliman}}, \bibinfo {author} {\bibfnamefont {M.}~\bibnamefont {Fan}}, \
  and\ \bibinfo {author} {\bibfnamefont {A.~M.}\ \bibnamefont {Jayich}},\
  }\href {\doibase 10.1103/PhysRevA.100.062512} {\bibfield  {journal} {\bibinfo
   {journal} {Phys. Rev. A}\ }\textbf {\bibinfo {volume} {100}},\ \bibinfo
  {pages} {062512} (\bibinfo {year} {2019})}\BibitemShut {NoStop}%
\bibitem [{\citenamefont {Miyake}\ \emph {et~al.}(2019)\citenamefont {Miyake},
  \citenamefont {Pisenti}, \citenamefont {Elgee}, \citenamefont {Sitaram},\
  and\ \citenamefont {Campbell}}]{Miyake2019}%
  \BibitemOpen
  \bibfield  {author} {\bibinfo {author} {\bibfnamefont {H.}~\bibnamefont
  {Miyake}}, \bibinfo {author} {\bibfnamefont {N.~C.}\ \bibnamefont {Pisenti}},
  \bibinfo {author} {\bibfnamefont {P.~K.}\ \bibnamefont {Elgee}}, \bibinfo
  {author} {\bibfnamefont {A.}~\bibnamefont {Sitaram}}, \ and\ \bibinfo
  {author} {\bibfnamefont {G.~K.}\ \bibnamefont {Campbell}},\ }\href {\doibase
  10.1103/PhysRevResearch.1.033113} {\bibfield  {journal} {\bibinfo  {journal}
  {Phys. Rev. Research}\ }\textbf {\bibinfo {volume} {1}},\ \bibinfo {pages}
  {033113} (\bibinfo {year} {2019})}\BibitemShut {NoStop}%
\bibitem [{\citenamefont {Meija}\ \emph {et~al.}(2016)\citenamefont {Meija}
  \emph {et~al.}}]{Meija2016}%
  \BibitemOpen
  \bibfield  {author} {\bibinfo {author} {\bibfnamefont {J.}~\bibnamefont
  {Meija}} \emph {et~al.},\ }\href@noop {} {\bibfield  {journal} {\bibinfo
  {journal} {Pure. Appl. Chem.}\ }\textbf {\bibinfo {volume} {88}},\ \bibinfo
  {pages} {293} (\bibinfo {year} {2016})}\BibitemShut {NoStop}%
\bibitem [{\citenamefont {Pitz}\ \emph {et~al.}(1990)\citenamefont {Pitz} \emph
  {et~al.}}]{Pitz1990}%
  \BibitemOpen
  \bibfield  {author} {\bibinfo {author} {\bibfnamefont {H.~H.}\ \bibnamefont
  {Pitz}} \emph {et~al.},\ }\href {\doibase
  https://doi.org/10.1016/0375-9474(90)90092-Z} {\bibfield  {journal} {\bibinfo
   {journal} {Nucl. Phys. A}\ }\textbf {\bibinfo {volume} {509}},\ \bibinfo
  {pages} {587 } (\bibinfo {year} {1990})}\BibitemShut {NoStop}%
\bibitem [{\citenamefont {Bhatt}\ \emph {et~al.}(2019)\citenamefont {Bhatt},
  \citenamefont {Kato},\ and\ \citenamefont {Vutha}}]{Bhatt2019}%
  \BibitemOpen
  \bibfield  {author} {\bibinfo {author} {\bibfnamefont {N.}~\bibnamefont
  {Bhatt}}, \bibinfo {author} {\bibfnamefont {K.}~\bibnamefont {Kato}}, \ and\
  \bibinfo {author} {\bibfnamefont {A.~C.}\ \bibnamefont {Vutha}},\ }\href
  {\doibase 10.1103/PhysRevA.100.013401} {\bibfield  {journal} {\bibinfo
  {journal} {Phys. Rev. A}\ }\textbf {\bibinfo {volume} {100}},\ \bibinfo
  {pages} {013401} (\bibinfo {year} {2019})}\BibitemShut {NoStop}%
\bibitem [{\citenamefont {King}\ \emph {et~al.}(1973)\citenamefont {King},
  \citenamefont {Steudel},\ and\ \citenamefont {Wilson}}]{King1973}%
  \BibitemOpen
  \bibfield  {author} {\bibinfo {author} {\bibfnamefont {W.~H.}\ \bibnamefont
  {King}}, \bibinfo {author} {\bibfnamefont {A.}~\bibnamefont {Steudel}}, \
  and\ \bibinfo {author} {\bibfnamefont {M.}~\bibnamefont {Wilson}},\
  }\href@noop {} {\bibfield  {journal} {\bibinfo  {journal} {Z. Phys.}\
  }\textbf {\bibinfo {volume} {265}},\ \bibinfo {pages} {207} (\bibinfo {year}
  {1973})}\BibitemShut {NoStop}%
\bibitem [{\citenamefont {Blaise}\ \emph {et~al.}(1984)\citenamefont {Blaise},
  \citenamefont {Wyart}, \citenamefont {Djerad},\ and\ \citenamefont
  {Ahmed}}]{Blaise1984}%
  \BibitemOpen
  \bibfield  {author} {\bibinfo {author} {\bibfnamefont {J.}~\bibnamefont
  {Blaise}}, \bibinfo {author} {\bibfnamefont {J.~F.}\ \bibnamefont {Wyart}},
  \bibinfo {author} {\bibfnamefont {M.~T.}\ \bibnamefont {Djerad}}, \ and\
  \bibinfo {author} {\bibfnamefont {Z.~B.}\ \bibnamefont {Ahmed}},\ }\href
  {\doibase 10.1088/0031-8949/29/2/005} {\bibfield  {journal} {\bibinfo
  {journal} {Phys. Scripta}\ }\textbf {\bibinfo {volume} {29}},\ \bibinfo
  {pages} {119} (\bibinfo {year} {1984})}\BibitemShut {NoStop}%
\bibitem [{\citenamefont {Nakhate}\ \emph {et~al.}(1997)\citenamefont
  {Nakhate}, \citenamefont {Afzal},\ and\ \citenamefont {Ahmad}}]{Nakhate1997}%
  \BibitemOpen
  \bibfield  {author} {\bibinfo {author} {\bibfnamefont {S.~G.}\ \bibnamefont
  {Nakhate}}, \bibinfo {author} {\bibfnamefont {S.~M.}\ \bibnamefont {Afzal}},
  \ and\ \bibinfo {author} {\bibfnamefont {S.~A.}\ \bibnamefont {Ahmad}},\
  }\href {\doibase 10.1007/s004600050334} {\bibfield  {journal} {\bibinfo
  {journal} {Z. Phys. D Atom. Mol. Cl.}\ }\textbf {\bibinfo {volume} {42}},\
  \bibinfo {pages} {71} (\bibinfo {year} {1997})}\BibitemShut {NoStop}%
\bibitem [{\citenamefont {Hongliang}\ \emph {et~al.}(1997)\citenamefont
  {Hongliang}, \citenamefont {Wei}, \citenamefont {Bin}, \citenamefont {Yong},
  \citenamefont {Dufei}, \citenamefont {Fuquan}, \citenamefont {Jiayong},\ and\
  \citenamefont {Fujia}}]{Hongliang1997}%
  \BibitemOpen
  \bibfield  {author} {\bibinfo {author} {\bibfnamefont {M.}~\bibnamefont
  {Hongliang}}, \bibinfo {author} {\bibfnamefont {S.}~\bibnamefont {Wei}},
  \bibinfo {author} {\bibfnamefont {Y.}~\bibnamefont {Bin}}, \bibinfo {author}
  {\bibfnamefont {L.}~\bibnamefont {Yong}}, \bibinfo {author} {\bibfnamefont
  {F.}~\bibnamefont {Dufei}}, \bibinfo {author} {\bibfnamefont
  {L.}~\bibnamefont {Fuquan}}, \bibinfo {author} {\bibfnamefont
  {T.}~\bibnamefont {Jiayong}}, \ and\ \bibinfo {author} {\bibfnamefont
  {Y.}~\bibnamefont {Fujia}},\ }\href {\doibase 10.1088/0953-4075/30/15/008}
  {\bibfield  {journal} {\bibinfo  {journal} {J. Phys. B}\ }\textbf {\bibinfo
  {volume} {30}},\ \bibinfo {pages} {3355} (\bibinfo {year}
  {1997})}\BibitemShut {NoStop}%
\bibitem [{\citenamefont {Koczorowski}\ \emph {et~al.}(2005)\citenamefont
  {Koczorowski}, \citenamefont {Stachowska}, \citenamefont {Furmann},
  \citenamefont {Stefa{\'{n}}ska}, \citenamefont {Jarosz}, \citenamefont
  {Krzykowski}, \citenamefont {Walaszyk}, \citenamefont {Szawio{\l}a},\ and\
  \citenamefont {Buczek}}]{Koczorowski2005}%
  \BibitemOpen
  \bibfield  {author} {\bibinfo {author} {\bibfnamefont {W.}~\bibnamefont
  {Koczorowski}}, \bibinfo {author} {\bibfnamefont {E.}~\bibnamefont
  {Stachowska}}, \bibinfo {author} {\bibfnamefont {B.}~\bibnamefont {Furmann}},
  \bibinfo {author} {\bibfnamefont {D.}~\bibnamefont {Stefa{\'{n}}ska}},
  \bibinfo {author} {\bibfnamefont {A.}~\bibnamefont {Jarosz}}, \bibinfo
  {author} {\bibfnamefont {A.}~\bibnamefont {Krzykowski}}, \bibinfo {author}
  {\bibfnamefont {A.}~\bibnamefont {Walaszyk}}, \bibinfo {author}
  {\bibfnamefont {G.}~\bibnamefont {Szawio{\l}a}}, \ and\ \bibinfo {author}
  {\bibfnamefont {A.}~\bibnamefont {Buczek}},\ }\href
  {https://linkinghub.elsevier.com/retrieve/pii/S058485470500025X} {\bibfield
  {journal} {\bibinfo  {journal} {Spectrochim. Acta. B}\ }\textbf {\bibinfo
  {volume} {60}},\ \bibinfo {pages} {447} (\bibinfo {year} {2005})}\BibitemShut
  {NoStop}%
\bibitem [{\citenamefont {Rosner}\ \emph {et~al.}(2005)\citenamefont {Rosner},
  \citenamefont {Masterman}, \citenamefont {Scholl},\ and\ \citenamefont
  {Holt}}]{Rosner2005}%
  \BibitemOpen
  \bibfield  {author} {\bibinfo {author} {\bibfnamefont {S.~D.}\ \bibnamefont
  {Rosner}}, \bibinfo {author} {\bibfnamefont {D.}~\bibnamefont {Masterman}},
  \bibinfo {author} {\bibfnamefont {T.~J.}\ \bibnamefont {Scholl}}, \ and\
  \bibinfo {author} {\bibfnamefont {R.~A.}\ \bibnamefont {Holt}},\ }\href
  {\doibase 10.1139/p05-029} {\bibfield  {journal} {\bibinfo  {journal} {Can.
  J. Phys.}\ }\textbf {\bibinfo {volume} {83}},\ \bibinfo {pages} {841}
  (\bibinfo {year} {2005})}\BibitemShut {NoStop}%
\bibitem [{\citenamefont {Jackson}\ \emph {et~al.}(2018)\citenamefont
  {Jackson}, \citenamefont {Sawaoka}, \citenamefont {Bhatt}, \citenamefont
  {Potnis},\ and\ \citenamefont {Vutha}}]{Jackson2018}%
  \BibitemOpen
  \bibfield  {author} {\bibinfo {author} {\bibfnamefont {S.}~\bibnamefont
  {Jackson}}, \bibinfo {author} {\bibfnamefont {H.}~\bibnamefont {Sawaoka}},
  \bibinfo {author} {\bibfnamefont {N.}~\bibnamefont {Bhatt}}, \bibinfo
  {author} {\bibfnamefont {S.}~\bibnamefont {Potnis}}, \ and\ \bibinfo {author}
  {\bibfnamefont {A.~C.}\ \bibnamefont {Vutha}},\ }\href {\doibase
  10.1063/1.5012000} {\bibfield  {journal} {\bibinfo  {journal} {Rev. Sci.
  Instrum.}\ }\textbf {\bibinfo {volume} {89}},\ \bibinfo {pages} {033109}
  (\bibinfo {year} {2018})}\BibitemShut {NoStop}%
\bibitem [{\citenamefont {Wang}\ \emph {et~al.}(2012)\citenamefont {Wang},
  \citenamefont {Audi}, \citenamefont {Wapstra}, \citenamefont {Kondev},
  \citenamefont {MacCormick}, \citenamefont {Xu},\ and\ \citenamefont
  {Pfeiffer}}]{Wang2012}%
  \BibitemOpen
  \bibfield  {author} {\bibinfo {author} {\bibfnamefont {M.}~\bibnamefont
  {Wang}}, \bibinfo {author} {\bibfnamefont {G.}~\bibnamefont {Audi}}, \bibinfo
  {author} {\bibfnamefont {A.}~\bibnamefont {Wapstra}}, \bibinfo {author}
  {\bibfnamefont {F.}~\bibnamefont {Kondev}}, \bibinfo {author} {\bibfnamefont
  {M.}~\bibnamefont {MacCormick}}, \bibinfo {author} {\bibfnamefont
  {X.}~\bibnamefont {Xu}}, \ and\ \bibinfo {author} {\bibfnamefont
  {B.}~\bibnamefont {Pfeiffer}},\ }\href
  {http://stacks.iop.org/1674-1137/36/i=12/a=003?key=crossref.f50105e9efa1d63fde56a731937fb5d1}
  {\bibfield  {journal} {\bibinfo  {journal} {Chinese Phys. C}\ }\textbf
  {\bibinfo {volume} {36}},\ \bibinfo {pages} {1603} (\bibinfo {year}
  {2012})}\BibitemShut {NoStop}%
\bibitem [{\citenamefont {Matei}\ \emph {et~al.}(2017)\citenamefont {Matei}
  \emph {et~al.}}]{Matei2017}%
  \BibitemOpen
  \bibfield  {author} {\bibinfo {author} {\bibfnamefont {D.~G.}\ \bibnamefont
  {Matei}} \emph {et~al.},\ }\href {\doibase 10.1103/PhysRevLett.118.263202}
  {\bibfield  {journal} {\bibinfo  {journal} {Phys. Rev. Lett.}\ }\textbf
  {\bibinfo {volume} {118}},\ \bibinfo {pages} {263202} (\bibinfo {year}
  {2017})}\BibitemShut {NoStop}%
\bibitem [{\citenamefont {Robinson}\ \emph {et~al.}(2019)\citenamefont
  {Robinson}, \citenamefont {Oelker}, \citenamefont {Milner}, \citenamefont
  {Zhang}, \citenamefont {Legero}, \citenamefont {Matei}, \citenamefont
  {Riehle}, \citenamefont {Sterr},\ and\ \citenamefont {Ye}}]{Robinson2019}%
  \BibitemOpen
  \bibfield  {author} {\bibinfo {author} {\bibfnamefont {J.~M.}\ \bibnamefont
  {Robinson}}, \bibinfo {author} {\bibfnamefont {E.}~\bibnamefont {Oelker}},
  \bibinfo {author} {\bibfnamefont {W.~R.}\ \bibnamefont {Milner}}, \bibinfo
  {author} {\bibfnamefont {W.}~\bibnamefont {Zhang}}, \bibinfo {author}
  {\bibfnamefont {T.}~\bibnamefont {Legero}}, \bibinfo {author} {\bibfnamefont
  {D.~G.}\ \bibnamefont {Matei}}, \bibinfo {author} {\bibfnamefont
  {F.}~\bibnamefont {Riehle}}, \bibinfo {author} {\bibfnamefont
  {U.}~\bibnamefont {Sterr}}, \ and\ \bibinfo {author} {\bibfnamefont
  {J.}~\bibnamefont {Ye}},\ }\href {\doibase 10.1364/OPTICA.6.000240}
  {\bibfield  {journal} {\bibinfo  {journal} {Optica}\ }\textbf {\bibinfo
  {volume} {6}},\ \bibinfo {pages} {240} (\bibinfo {year} {2019})}\BibitemShut
  {NoStop}%
\bibitem [{\citenamefont {Kramida}\ \emph {et~al.}(2019)\citenamefont
  {Kramida}, \citenamefont {{Yu.~Ralchenko}}, \citenamefont {Reader},\ and\
  \citenamefont {{and NIST ASD Team}}}]{NIST_ASD}%
  \BibitemOpen
  \bibfield  {author} {\bibinfo {author} {\bibfnamefont {A.}~\bibnamefont
  {Kramida}}, \bibinfo {author} {\bibnamefont {{Yu.~Ralchenko}}}, \bibinfo
  {author} {\bibfnamefont {J.}~\bibnamefont {Reader}}, \ and\ \bibinfo {author}
  {\bibnamefont {{and NIST ASD Team}}},\ }\href@noop {} {}\bibinfo
  {howpublished} {{NIST Atomic Spectra Database (ver. 5.7.1), [Online].
  Available: {\tt{https://physics.nist.gov/asd}} [2020, February 11]. National
  Institute of Standards and Technology, Gaithersburg, MD.}} (\bibinfo {year}
  {2019})\BibitemShut {NoStop}%
\end{thebibliography}%

\end{document}